\documentclass[seceq,supplement]{ptptex}

\usepackage{graphicx}

\title{Replacing the Breit-Wigner amplitude by the complex delta 
function to describe resonances}

\author{Rafael \textsc{de la Madrid}}

\inst{Department of Physics, Lamar University,
Beaumont, TX 77710 \\
\small{E-mail: \texttt{rafael.delamadrid@lamar.edu}}}

\abst{\noindent Whenever the Breit-Wigner amplitude appears in a calculation,
there are many instances (e.g., Fermi's two-level system and the
Weisskopf-Wigner approximation) where energy integrations are extended from the
scattering spectrum of the Hamiltonian to the whole real line. Such extensions 
are performed in order to obtain a desirable, causal result. In this paper, 
we recall several of those instances and show that substituting the 
Breit-Wigner amplitude by the complex delta function allows us to
recover such desirable results without having to extend
energy integrations outside of the scattering spectrum.}

\begin{document}

\maketitle
















\section{Introduction}
\setcounter{equation}{0}
\label{sec:Intro}

As is well known, the decay/resonant amplitude of a resonance cannot 
exactly coincide with the Breit-Wigner amplitude. One reason is that the 
Breit-Wigner 
amplitude yields the exponential decay law only when it is defined over the
whole of the energy real line $(-\infty , \infty )$ rather than just over the
scattering spectrum. Because in quantum mechanics 
the scattering spectrum has a lower bound, the Breit-Wigner amplitude would 
yield the exponential decay law only if it was defined also at energies that 
do not belong to the scattering spectrum. 

Another reason why the Breit-Wigner amplitude cannot exactly coincide with the 
resonant amplitude is that the energy (i.e., spectral) representation of the
Gamow state is given by the complex delta function rather than by the
Breit-Wigner amplitude. Because the Gamow state is the natural wave function
of a resonance, the exact resonant amplitude is given by the complex delta 
function.

Even though it is well known that it cannot exactly coincide with the 
resonant amplitude, the Breit-Wigner amplitude is often used
to describe the decay of unstable systems. Two classic examples are Fermi's 
two-level system and the Weisskopf-Wigner approximation. However, whenever
it is used in such examples, the Breit-Wigner amplitude is 
extended from the scattering spectrum to the whole real line of energies
in order to obtain some desirable, causal results.

Because the exact resonant amplitude is given by the complex delta function,
one may wonder if such desirable results can be recovered by way of the
complex delta function without using the approximation of extending energy
integrations to the whole real line. The purpose of this
paper is to show how this can be done.

In Secs.~\ref{sec:Fermi}, \ref{sec:other} and~\ref{sec:WWAPRSO}, we recall,
respectively, the main features of Fermi's two-level system, a standard
treatment of unstable states, and the Weisskopf-Wigner approximation.
In Sec.~\ref{sec:subs}, we explain why the complex delta function gives us
the same results as the Breit-Wigner amplitude without extending energy 
integrations outside of the scattering spectrum. In Sec.~\ref{sec:con}, we state
our conclusions.

\section{Fermi's two level system}
\setcounter{equation}{0}
\label{sec:Fermi}

\begin{figure}[ht]
\hskip4cm \includegraphics[width=8.5cm]{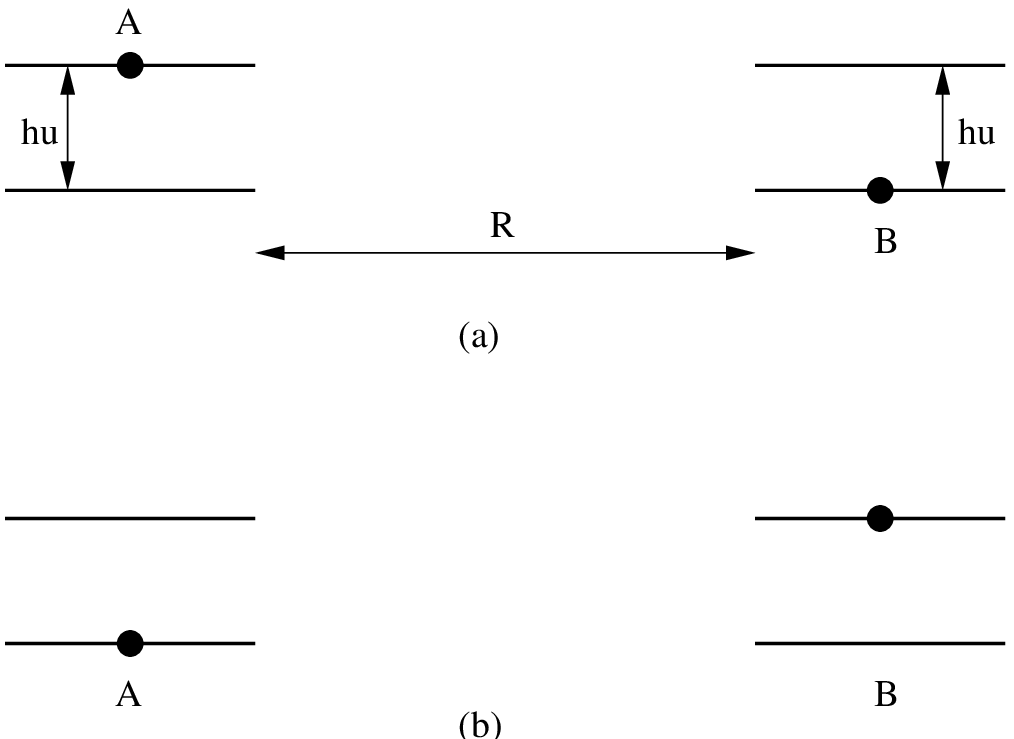}
\caption{Schematic representation of the (a) initial and (b) final states of
Fermi's two-level system.}
\label{states}
\end{figure}

\begin{figure}[ht]
\hskip4cm \includegraphics[width=8.5cm]{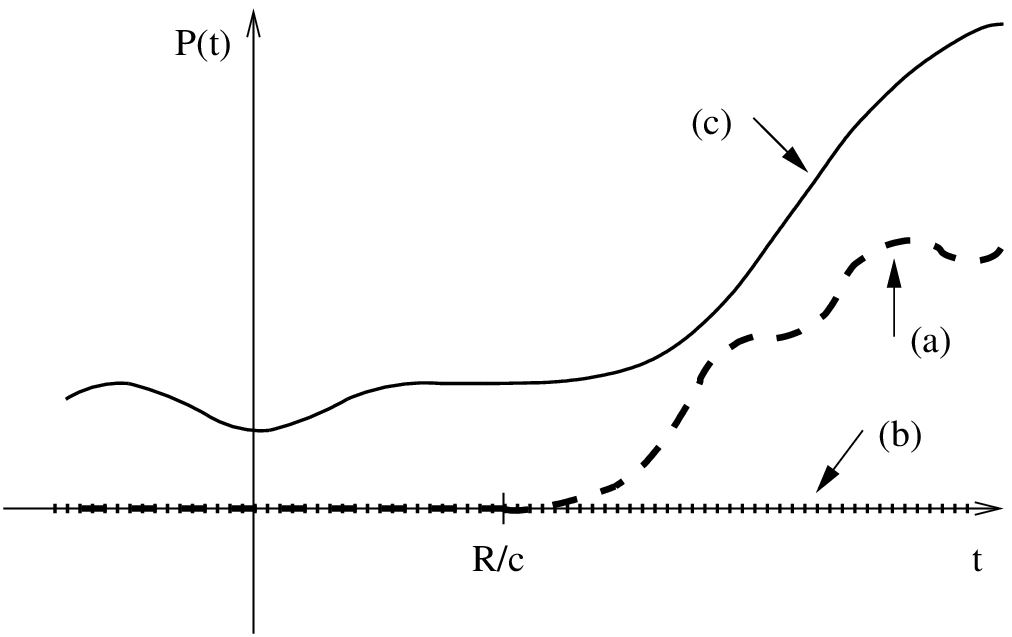}
\caption{Schematic representation of (a) Fermi's causal result, (b) first 
outcome of Hegerfeldt's theorem, and (c) second outcome of Hegerfeldt's 
theorem.}
\label{outcome}
\end{figure}

In 1932, Fermi constructed a simple model to check whether Quantum Mechanics 
is compatible with Einstein causality~\cite{FERMI32}. He considered a pair
of two atoms A and B separated by a distance $R$. The states of each atom 
form a two-level
system (see Fig.~\ref{states}a). The energy gap of each two-level system is 
$hu$ (see Fig.~\ref{states}a). The initial state is such that atom A is in 
the excited state, whereas atom B is in the ground
state (see Fig.~\ref{states}a). When atom A decays to its ground state,
it emits a photon of energy $hu$. This photon may eventually hit 
atom~B, causing atom~B to reach the excited state. The final 
state is such that atom A~is in the ground 
state, whereas atom~B is in the excited state (see Fig.~\ref{states}b). Fermi 
then calculated the probability ${\cal P}_{i\to f}(t)$ of going from 
the initial state of Fig.~\ref{states}a to the final state of 
Fig.~\ref{states}b. According to Einstein causality, ${\cal P}_{i\to f}(t)$ 
should be zero for any instant $t$ less than $R/c$, i.e., for any $t$ 
less 
than what it takes the photon to go from atom A to atom B (see 
Fig.~\ref{outcome}a). This is the result that Fermi obtained~\cite{FERMI32}.  

About the same time Fermi proposed this model, von Neumann published his
book on the mathematical foundations of Quantum 
Mechanics~\cite{vonNeuman}. According to von Neumann, the energy observable 
is represented by a linear, self-adjoint operator, called Hamiltonian, that 
acts on a Hilbert space. The spectrum of the Hamiltonian, which is 
identified with the physical spectrum, should be bounded from below
(i.e., semibounded).  

In 1966, Shirokov~\cite{Shirokov66} pointed out that, in order to obtain the 
result of Fig.~\ref{outcome}a, Fermi had approximated an integral over 
positive energies (i.e., over the scattering spectrum) by an integral over the 
full energy real line $(-\infty ,\infty )$. Such integral involves the
Breit-Wigner amplitude. This approximation is 
{\it crucial} to Fermi's calculation: if the integral is performed over the
scattering spectrum, then the causal result of Fig.~\ref{outcome}a does not 
hold~\cite{Shirokov66}. In fact, in 1994 Hegerfeldt~\cite{Hegerfeldt94} 
showed, in a model independent manner, that the problem pointed out by Shirokov 
within Fermi's system is quite general: the semiboundedness of the Hamiltonian 
leads to conflicts with causality. More precisely, according to
Hegerfeldt's theorem, Quantum Mechanics predicts that
either atom A never decays (see Fig.~\ref{outcome}b), or else there is 
a non-zero probability that atom~B reaches the excited state before
the photon from atom A can possibly arrive at atom~B (see Fig.~\ref{outcome}c).

\section{Unstable states}
\setcounter{equation}{0}
\label{sec:other}

Approximations similar to Fermi's approximation can be found in standard 
textbooks dealing with unstable states. For example, in Sec.~13.d of 
Ref.~\cite{TAYLOR},
Taylor uses such kind of approximation when dealing with the
decay of a resonant state. More precisely, Taylor's equation~(13.3) reads
as
\begin{equation}
           \psi _{\rm sc}({\bf x},t)= {\rm constant}\
               \Gamma Y_l^0(\hat{\rm x})\phi_l(E_{\rm R})
               \frac{e^{i(p_{\rm R}r-E_{\rm R}t)}}{p_{\rm R}^{1/2}r}
               \times \int_0^{\infty}dE 
                 \frac{e^{i(E-E_{\rm R})(t-r/v_{\rm R})}}{E-E_{\rm R}+i\Gamma /2} \,  ,
\end{equation}
where $z_{\rm R}= E_{\rm R}-i\Gamma /2$ is the complex resonant energy. Taylor 
then continues by saying that ``the integral can be
extended to $-\infty$ without significantly affecting its value.'' After
such extension, Taylor obtains the following desirable result:
\begin{equation}
        |\psi _{\rm sc}({\bf x},t)|^2= 2 \pi m 
               \Gamma^2 |Y_l^0(\hat{\rm x})|^2 |\phi_l(E_{\rm R})|^2
                \frac{e^{-\Gamma(t-r/v_{\rm R})}}{p_{\rm R}r^2}
                \theta \left( t -\frac{r}{v_{\rm R}}\right) .
          \label{psideca}
\end{equation}
Equation~(\ref{psideca}) implies that the decay of a resonant state
follows the exponential decay law in a causal manner. Clearly, 
Eq.~(\ref{psideca}) does not hold exactly when the integration
is done over the scattering spectrum, just like causality is not preserved
in Fermi's two-level system when the integration is done over the
scattering spectrum. 

In Ref.~\cite{BALLENTINE}, Ballentine treats the decay of a resonance
in a similar way to Taylor. Ballentine also extends an energy
integral to the whole real line (see Eq.~(16.120) of Ref.~\cite{BALLENTINE})
in order to obtain a desirable, causal result.

\section{Weisskopf-Wigner approximation}
\setcounter{equation}{0}
\label{sec:WWAPRSO}

In quantum mechanics, the approximation of extending the range of the
Breit-Wigner amplitude to the whole real line is often referred to as
the Weisskopf-Wigner approximation. Such approximation is used in many
calculations. For example, in Ref.~\cite{SCULLY}, Scully and Zubairy
calculate the following amplitude for the first-order correlation function:
\begin{equation}
    \langle 0|E^{(+)}({\bf r},t)|\gamma _0\rangle =
    \frac{ic {\cal P}_{ab} \sin \eta}{8\pi ^2 \epsilon _0 \Delta r}
    \times \int_0^{\infty}dk k^2(e^{ik\Delta r}- e^{-ik\Delta r})
      \frac{e^{-i\nu _kt}}{(\nu _k -\omega )+i\Gamma /2} \, .
              \label{lrigkdkt}
\end{equation}
Then, Scully and Zubairy extend the range of the integral to the 
whole real line and obtain a desirable causal result for the first-order 
correlation function:
\begin{equation}
   G^{(1)}({\bf r},{\bf r};t,t)= 
      |\langle 0|E^{(+)}({\bf r},t)|\gamma _0\rangle|^2 =
  \frac{ |{\cal E}_0|^2}{|{\bf r}-{\bf r}_0|^2}
    \theta (t- \frac{|{\bf r}-{\bf r}_0|}{c}) 
     e^{-\Gamma (t-|{\bf r}-{\bf r}_0|/c)} \, .
\end{equation}
As in the above examples, this result cannot be obtained unless the
range of the frequency (energy) integration in Eq.~(\ref{lrigkdkt})
is extended to the whole real line.

\section{Substituting the Breit-Wigner amplitude by the complex
delta function}
\setcounter{equation}{0}
\label{sec:subs}

From the above examples, we have seen that, whenever we describe the
decay of an unstable state by the Breit-Wigner amplitude, we arrive
at an integral of the form
\begin{equation}
      \int_0^{\infty}dE \, e^{-iEt} \frac{f(E)}{E-z_{\rm R}} \, ,
           \label{ldlds}
\end{equation}
where $f(E)$ is an analytic function of $E$, and 
$z_{\rm R}=E_{\rm R}-i\Gamma /2$ is the resonant energy. By assuming that the
extension of the integral to the whole real line
makes little error, one gets
\begin{equation}
      \int_{-\infty}^{\infty}dE \, e^{-iEt} f(E) \frac{1}{E-z_{\rm R}} = 
        \frac{2 \pi}{i} 
       f(z_{\rm R}) e^{-iE_{\rm R}t}e^{-\Gamma t/2} \, , \qquad t>0 \, .
          \label{desidkdkd}
\end{equation}
Equations similar 
to~(\ref{desidkdkd}) are widely used in the literature on resonances 
(see e.g.~review~\cite{ROSAS}). 

Clearly, the desirable result~(\ref{desidkdkd}) is obtained by using the
approximation of extending the energy integration to the whole real 
line. It seems therefore pertinent to try to 
recover~(\ref{desidkdkd}) as an exact result. In order to do so, we are 
going to substitute the Breit-Wigner amplitude by the complex delta function.

The complex delta function was introduced by Nakanishi~\cite{NAKANISHI} 
to describe resonances in the Lee model~\cite{LEE}, and it has been used by 
a number of authors (see e.g.~Refs.~\cite{GONZALO00,GONZALO1,GONZALO2} and 
references therein). As shown in Ref.~\cite{NPA}, describing resonances by 
means of the complex 
delta function is the same as describing resonances by means of the 
Gamow state~\cite{GAMOW,SIEGERT,PEIERLS,HUMBLET,ZELDOVICH,BERGGREN,GASTON,
BERGGREN78,SUDARSHAN,MONDRAGON83,CURUTCHET,BL,LIND,BERGGREN96,BOLLINI,
FERREIRA,BETAN,MICHEL1,AJP02,KAPUSCIK1,MONDRAGON03,MICHEL2,KAPUSCIK2,05CJP,
MICHEL3,MICHEL4,MICHEL5,URRIES,MICHEL6,COSTIN,ROSAS22,TOMIO,MICHEL7,
HATANO,HUANG}. We recall that the Gamow states are eigenfunctions of the
Hamiltonian subject to purely boundary conditions. The eigenvalue of the
Gamow state is also a pole of the $S$ matrix. The 
resonant amplitude associated with the Gamow states is given
by the complex delta function, and the Breit-Wigner amplitude is just
an approximate resonant amplitude that is valid whenever we neglect the lower 
bound of the energy.\footnote{One may argue that the Breit-Winger amplitude is 
an approximation also because the exact formula corresponding to
the denominator $E-z_{\rm R}$ is a much more complicated function of $E$ that
includes a self-energy contribution.} 

Mathematically, the complex delta function is a distribution that associates,
with a test function $g$, the value of such function at $z=z_{\rm R}$:
\begin{equation}
      \int_{0}^{\infty}dE \, g(E) \delta (E-z_{\rm R}) = g(z_{\rm R}) \, .
          \label{compdfdef}
\end{equation}
Now, if the resonant amplitude is given by the complex delta function
(rather than by the Breit-Wigner amplitude
$\frac{1}{E-z_{\rm R}}$), and if the scattering spectrum is the
positive real line (rather than the whole real line), Eq.~(\ref{ldlds}) 
should be written as
\begin{equation}
      \int_{0}^{\infty}dE \, e^{-iEt} f(E) \delta (E-z_{\rm R})  \, .
           \label{deltaintpos}
\end{equation}
By combining Eqs.~(\ref{compdfdef}) and~(\ref{deltaintpos}), we obtain
\begin{equation}
      \int_{0}^{\infty}dE \, e^{-iEt} f(E) \delta (E-z_{\rm R}) =
            f(z_{\rm R}) e^{-iE_{\rm R}t}e^{-\Gamma t/2}  \, ,
           \label{aadeltaintpos}
\end{equation}
which, up to a numerical factor, coincides with the 
desirable result~(\ref{desidkdkd}). In addition, since the time 
evolution of the complex delta function is defined only for 
$t>0$ (see Refs.~\cite{GONZALO00,GONZALO1,GONZALO2}), Eq.~(\ref{aadeltaintpos})
is valid only for $t>0$:
\begin{equation}
      \int_{0}^{\infty}dE \, e^{-iEt} f(E) \delta (E-z_{\rm R}) =
            f(z_{\rm R}) e^{-iE_{\rm R}t}e^{-\Gamma t/2}  \, , \qquad t>0 \, .
           \label{aadeltaintposaa}
\end{equation}
Thus, instead of using the Breit-Wigner amplitude and integrating over the 
whole real line as in Eq.~(\ref{desidkdkd}), we can integrate over 
the scattering spectrum and use the complex delta function as in 
Eq.~(\ref{aadeltaintposaa}) to obtain the same result.

\section{Conclusion}
\setcounter{equation}{0}
\label{sec:con}

In Quantum Mechanics, the combination of two approximations --the approximation 
of describing the decay of an unstable state by means of the 
Breit-Wigner amplitude, and the approximation of 
extending the Breit-Wigner amplitude to the whole real line-- yields
desirable, causal results for the decay of a resonance. In this paper, we 
have seen that if we replace the Breit-Wigner amplitude by the complex delta 
function, it is possible
to recover such desirable results without the need to extend any energy
integration outside of the physical scattering spectrum. This result provides
another argument in favor of seeing the complex delta function as
the exact resonant amplitude, and the Breit-Wigner amplitude as an
approximate resonant amplitude that is valid whenever we can neglect
the lower bound of the energy\cite{NPA}. This results also shows that 
what Fermi's, Weisskopf-Wigner's and similar approximations do is
basically to neglect the effect of the lower bound of the energy.

\vskip1cm

{\it Acknowledgment}. The author wishes to thank the organizers
of the YKIS2009 workshop for their invitation and warm hospitality. The
author also wishes to thank the participants for lively, enlightening
discussions, especially Profs.~Tomio Petrosky, Naomichi Hatano, Gonzalo
Ordo\~nez, and Buang Ann Tay.

\newpage

\begin{thebibliography}{99}

\bibitem{FERMI32} E.~Fermi, Rev.~Mod.~Phys.~{\bf 4}, (1932) 87.

\bibitem{vonNeuman} J.\ von Neumann, {\it Mathematische Grundlagen der
Quantentheorie} (Springer, Berlin) 1931.

\bibitem{Shirokov66} M.I.~Shirokov, Yad.~Fiz.~{\bf 4}, (1966) 1077 
[Sov.~J.~Nucl.~Phys.~{\bf 4}, (1967) 744].

\bibitem{Hegerfeldt94} G.C.~Hegerfeldt, Phys.~Rev.~Lett.~{\bf 72}, 
(1994) 596.

\bibitem{TAYLOR} J.R.~Taylor, {\it Scattering Theory}, John Wiley \&
Sons (1972)

\bibitem{BALLENTINE} L.E.~Ballentine, {\it Quantum Mechanics: A modern
development}, World Scientific (2000).

\bibitem{SCULLY} M.O.~Scully, M.S.~Zubairy, {\it Quantum Optics}, 
Cambridge University Press (1997)

\bibitem{ROSAS} O.~Rosas-Ortiz, N.~Fernandez-Garcia, Sara Cruz y Cruz,
AIP Conference Proceedings {\bf 1077}, (2008) 31; {\sf arXiv:0902.4061} 

\bibitem{NAKANISHI} N.~Nakanishi, Prog.~Theo.~Phys.~{\bf 19}, (1958) 607.

\bibitem{LEE} T.D.~Lee, Phys.~Rev.~{\bf 95}, (1954) 1329.

\bibitem{GONZALO00} T.~Petrosky, G.~Ordo\~nez, and I.~Prigogine, 
Phys.~Rev.~A, {\bf 62}, (2000) 042106. 

\bibitem{GONZALO1} G.~Ordo\~nez, T.~Petrosky, and I.~Prigogine, 
Phys.~Rev.~A, {\bf 63}, (2001) 052106.

\bibitem{GONZALO2} T.~Petrosky, G.~Ordo\~nez, and I.~Prigogine, 
Phys.~Rev.~A, {\bf 64}, (2001) 062101.

\bibitem{NPA} R.~de la Madrid, 	Nucl.~Phys.~A~{\bf 812}, (2008) 13;
{\sf arXiv:0810.0876}

\bibitem{GAMOW}  G.~Gamow, Z.~Phys.~{\bf 51}, (1928) 204.

\bibitem{SIEGERT} A.F.J.~Siegert, Phys.~Rev.~{\bf 56}, (1939) 750.

\bibitem{PEIERLS} R.E.~Peierls, Proc.~R.~Soc.~(London) A {\bf 253}, (1959) 16.

\bibitem{HUMBLET} J.~Humblet, L.~Rosenfeld, 
Nucl.~Phys.~{\bf 26}, (1961) 529.

\bibitem{ZELDOVICH} Ya.B.~Zeldovich, Sov.~Phys.~JETP~{\bf 12}, (1961) 542.

\bibitem{BERGGREN} T.~Berggren, Nucl.~Phys.~A~{\bf 109}, (1968) 265.

\bibitem{GASTON} G.~Garc\'\i a-Calder\'on, R.~Peierls,
Nucl.~Phys.~A~{\bf 265}, (1976) 443.

\bibitem{BERGGREN78} T.~Berggren, Phys.~Lett.~B~{\bf 73}, (1978) 389.

\bibitem{SUDARSHAN} G.~Parravicini, V.~Gorini, E.C.G.~Sudarshan, 
J.~Math.~Phys.~{\bf 21}, (1980) 2208.

\bibitem{MONDRAGON83} E.~Hern\'andez, A.~Mondrag\'on, 
Phys.~Rev.~C~{\bf 29}, (1984) 722.

\bibitem{CURUTCHET} P.~Curutchet, T.~Vertse, R.J.~Liotta, 
Phys.~Rev.~C~{\bf 39}, (1989) 1020.

\bibitem{BL} T.~Berggren, P.~Lind, Phys.~Rev.~C~{\bf 47}, (1993) 768.

\bibitem{LIND} P.~Lind, Phys.~Rev.~C~{\bf 47}, (1993) 1903.

\bibitem{BERGGREN96} T.~Berggren, Phys.~Lett.~B~{\bf 373}, (1996) 1.

\bibitem{BOLLINI} C.G.~Bollini, O.~Civitarese, A.L.~De Paoli, M.C.~Rocca,
Phys.~Lett.~B{\bf 382}, (1996) 205.

\bibitem{FERREIRA} L.S.~Ferreira, E.~Maglione, R.J.~Liotta, 
Phys.~Rev.~Lett.~{\bf 78}, (1997) 1640.

\bibitem{BETAN} R.~Id Betan, R.J.~Liotta, N.~Sandulescu, T.~Vertse,
Phys.\ Rev.\ Lett.\ {\bf 89}, (2002) 042501.

\bibitem{MICHEL1} N.~Michel, W.~Nazarewicz, M.~Ploszajczak, K.~Bennaceur,
Phys.~Rev.~Lett.~{\bf 89}, (2002) 042502.

\bibitem{AJP02} R.~de la Madrid, M.~Gadella, Am.~J.~Phys.~{\bf 70}, 
(2002) 626.

\bibitem{KAPUSCIK1} E.~Kapuscik, P.~Szczeszek, Czech.\ J.\ Phys.\ {\bf 53}, 
(2003) 1053.

\bibitem{MONDRAGON03} E.~Hern\'andez, A.~J\'auregui, A.~Mondrag\'on, 
Phys.~Rev.~A~{\bf 67}, (2003) 022721.

\bibitem{MICHEL2} N.~Michel, W.~Nazarewicz, M.~Ploszajczak, J.~Okolowicz,
Phys.~Rev.~C~{\bf 67}, (2003) 054311.

\bibitem{KAPUSCIK2} E.~Kapuscik, P.~Szczeszek, Found.\ Phys.\ Lett.\ {\bf 18}, 
(2005) 573.

\bibitem{05CJP} R.~de la Madrid, G.~Garcia-Calderon, J.G.~Muga,
Czech.\ J.\ Phys.~{\bf 55}, (2005) 1141; {\sf quant-ph/0512242}.

\bibitem{MICHEL3} N.~Michel, W.~Nazarewicz, M.~Ploszajczak, J.~Rotureau,
Phys.~Rev.~C~{\bf 74}, (2006) 054305; {\sf nucl-th/0609016}.

\bibitem{MICHEL4} J.~Rotureau, N.~Michel, W.~Nazarewicz, M.~Ploszajczak, 
J.~Dukelsky, Phys.~Rev.~Lett.~{\bf 97}, (2006) 110603; {\sf nucl-th/0603021}.

\bibitem{MICHEL5} N.~Michel, W.~Nazarewicz, M.~Ploszajczak,
Phys.~Rev.~C~{\bf 75}, (2007) 031301; {\sf nucl-th/0702021}.

\bibitem{URRIES} J.~Julve, F.J.~de Urr{\'\i}es, J.~Phys.~A: 
Math.~Theor.~{\bf 41}, (2008) 304010; {\sf quant-ph/0701213}.


\bibitem{MICHEL6} N.~Michel, W.~Nazarewicz, M.~Ploszajczak,
Nuc.~Phys.~A~{\bf 794}, (2007) 29; {\sf arXiv:0707.0767}.


\bibitem{COSTIN} O.~Costin, J.L.~Lebowitz, C.~Stucchio,
Rev.~Math.~Phys.~{\bf 20}, (2008) 835; {\sf math-ph/0609069}.


\bibitem{ROSAS22} N.~Fernandez-Garcia, O.~Rosas-Ortiz, Ann.~Physics~{\bf 323},
(2008) 1397; {\sf arXiv:0810.5597}.


\bibitem{TOMIO} N.~Hatano, K.~Sasada, H.~Nakamura, T.~Petrosky, 
Prog.~Theo.~Phys.~{\bf 119}, (2008) 187; {\sf arXiv:0705.1388}.

\bibitem{MICHEL7} N.~Michel, W.~Nazarewicz, M.~Ploszajczak, T.~Vertse, 
J.~Phys.~G: Nucl.~Part.~Phys.~{\bf 36}, (2009) 013101; {\sf arXiv:0810.2728}.

\bibitem{HATANO} N.~Hatano, T.~Kawamoto, J.~Feinberg, Pramana~{\bf 73}, 
(2009) 553; {\sf arXiv:0904.1044}.

\bibitem{HUANG} M.~Huang, J.~Stat.~Phys.~{\bf 137}, (2009) 569;
{\sf arXiv:0904.4040}.



\end{thebibliography}
\end{document}